\RequirePackage{fix-cm}
\documentclass[nologo,11pt,a4paper]{ETHpaper}
\usepackage{graphicx}
\usepackage{hyperref} %
\usepackage{bm} %
\usepackage{caption} %
\usepackage{multirow} %
\usepackage{natbib} %

\usepackage{booktabs}

\usepackage{color}
\usepackage[usenames,dvipsnames]{xcolor}
\definecolor{eth_red}{HTML}{A8322D}
\definecolor{eth_green}{HTML}{3C5A0F}
\definecolor{eth_blue}{HTML}{1F407A}

\usepackage[textsize=tiny]{todonotes}

\usepackage{xargs}
\reversemarginpar

\title{When standard network measures fail to rank journals: A theoretical and empirical analysis}
\titlealternative{When standard network measures fail to rank journals}
\author{Giacomo Vaccario, Luca Verginer}
\authoralternative{G. Vaccario,  L. Verginer}
\address{Chair of Systems Design, ETH Zurich, Switzerland\\
  \url{www.sg.ethz.ch}}
\www{\url{http://www.sg.ethz.ch}}
\begin{document}

\maketitle

\begin{abstract}
Journal rankings are widely used and are often based on citation
data in combination with a network perspective.
We argue that some of these network-based rankings can produce misleading results.
From a theoretical point of view, we show that the standard network modelling approach of citation data at the journal level (i.e., the projection of paper citations onto journals) introduces fictitious relations among journals.
To overcome this problem, we propose a citation path perspective, and empirically show that rankings based on the network and the citation path perspective are very different.
Based on our theoretical and empirical analysis, we highlight the limitations of standard network metrics, and propose a method to overcome these limitations and compute journal rankings.
\end{abstract}

\section{Introduction}\label{sec:intro}

Bibliometricians and scientometricians often use citation-based indicators to rank and evaluate articles, journals, and authors in academic publishing \citep{owens2013research,hicks2015bibliometrics}.
The impact factor and h-index are among the most widely used indicators to assess journals~\citep{garfield1964science,hirsch2005index,braun2006hirsch}.
These indicators are \textit{local} in the sense that they are based on the number of citations received by a given article, author, or journal within a given time period.
More sophisticated indicators have been developed using citation data and network analysis, such as the journal influence measure by \citet{pinski1976citation}, a precursor to PageRank~\citep{brin1998anatomy}, and the SCImago Journal Rank (SJR) indicator \citep{guerrero2012further}.
These indicators are based on eigenvector centralities and rely on \textit{non-local} information.
The rationale to use non-local information is to give more weight to citations from papers that are well-cited.

The assumption at the core of both local and non-local indicators is that the citing paper is influenced by the cited one.
This assumption is motivated in two ways, namely (1) by {knowledge flow} and (2) and the {allocation of scientific credit}.
Specifically, it is assumed that knowledge flows in the opposite direction of citations.
Thus, a paper receiving many citations contains knowledge that is often reused to create new knowledge, i.e., new papers.
Similarly, authors endorse each other by citing their works, and hence, citations proxy credit allocation.
Non-local indicators also rely on the {path transitivity} assumption, i.e., given a network all sequences of links represent a possible path.
For example, given two paper citations $(c \to b)$ and $(b \to a)$,
the transitivity assumption implies that there is a path $(c \to b \to a)$, and hence, paper $c$ may influence paper $a$ via $b$.
In other words, there is a possible causal connection between the three papers.
We argue that the projection of citations among papers onto journals violates this transitivity assumption, and that the causal connection is lost.
We show that this violation affects journal rankings derived from non-local indicators.

The path transitivity assumption is not justified in the citation network at the journal level for two reasons.
{First}, the {journal aggregation} of the citation links may violate the path transitivity assumption.
Given two consecutive links between journal $A$, $B$ and $C$,
we do not know if the paper in $B$ cited by the paper in $A$ is also the paper citing the paper in $C$.
Hence, we do not know if there was any influence from $A$ to $C$ via $B$.
Path transitivity would instead incorrectly imply the presence of a path between $A$ to $C$.
{Second}, the {time aggregation} of citation links also violates path transitivity since we lose the ordering of citation events.
In order words, when aggregating citations of papers published at different times, one erroneously assumes that younger papers can influence older ones.

In the present work, we study the effect of violating the path transitivity assumption in general.
Note that our argumentation is valid for the knowledge flow and the scientific credit allocation perspectives.
For this reason, we will use the term \textit{fictitious influence} to refer to both.

The remainder of this paper is structured as follow.
In Section~\ref{sec:literature_review}, we briefly review the usage of journal rankings and recent findings in network science highlighting the importance of the path transitivity assumption.
Section~\ref{sec:comp-paths-and-nets} clarifies the pitfalls in projecting paper citations onto journals. %
In Section~\ref{sec:recon-know-diff}, we show empirically how journal rankings are biased by fictitious influence.
Finally, in Section~\ref{sec:chap4-con}, we summarize and discuss our results.

\section{Literature Review}\label{sec:literature_review}

Scientometricians and bibliometricians traditionally use citation analysis to develop quantitative indicators.
These indicators are obtained by identifying the properties of documents through their cross-referencing.
One example is the commonly used impact factor~\citep{garfield1964science}.
It captures the influence of journals by computing the average number of citations received by papers published in them.
More sophisticated indicators have been developed by combining citation with network analysis.
Specifically, practitioners have used this analysis by constructing a citation network at the journal level.
In this network, journals are nodes, and links are citations among papers published in them.
Network measures, such as eigenvector and betweenness centralities, have been proposed as indicators to determine journal influence~\citep{pinski1976citation,guerrero2012further}
and their interdisciplinarity~\citep{leydesdorff2007betweenness, leydesdorff2018betweenness}.
Moreover, such measures have been used to quantify the influence of authors~\citep{radicchi2009diffusion} and papers \citep{chen2007finding,zhou2016ranking}.

As mentioned in the introduction, the use of citation data is motivated by the credit allocation mechanism.
In other words, we assume that when an author cites a paper, he endorses the authors of the cited paper.
When projecting citations onto journals, we implicitly assume the same, namely that citation links among journals capture credit allocation from one journal to the other.
Additionally, most network measures rely on the \textit{path transitivity assumption}. When inferring (from data) the existence of links from $A$ to $B$ and $B$ to $C$, we automatically permit a path of length two from $A$ to $C$ via $B$.
Specifically, practitioners rely implicitly on this assumption to construct paths from citation links at the journal level.
These paths represent possible flows of knowledge between journals and have been used to compute journals' similarity~\citep{small1977journal}, journal influence~\citep{pinski1976citation}, and journal interdisciplinarity~\citep{leydesdorff2007betweenness, leydesdorff2018betweenness}.

Despite the proliferation and wide usage of citation-based indicators, they are also criticized.
A first concern arises from the fact that the citation practices vary across scientific fields~\citep{schubert1986relative,bornmann2008citation,radicchi2008universality}.
These differences introduce biases in citation-based indicators that cannot be easily overcome~\citep{albarran2011skewness, waltman2012universality,vaccario2017quantifying}.
A second concern relates to the fact that publications are increasingly written by multiple co-authors.
Various works have shown that co-authorship and the number of citations are deeply intertwined~\citep{persson2004inflationary,sarigol2014predicting,parolo2015attention, nanumyan2019citations}.
Further concerns about using citation and bibliographic data come from the results on how editorial biases relate to social factors, such as previous co-authorship~\citep{sarigol2017quantifying, dondio2019invisible} and citation remuneration \citep{petersen2019megajournal}.
These findings with many others questioned the objectivity of citation-based indicators.

Recent advances in network theory have also raised concerns about the naive applications of network analytic tools to complex data~\citep{butts2009revisiting, zweig2011good, borgatti2020three}.
In particular,
\cite{butts2009revisiting} stresses the importance to correctly match the unit and purpose of the analysis with the appropriate network representation.
These concerns, we argue, are also valid when one applies network measures to rank journals using paper citations.
To do this, one moves the unit of analysis from papers to journals without a full understanding of the implications.
Moreover, \citet{mariani2015ranking} show how PageRank fails to identify significant nodes in time-evolving networks.
This problem particularly applies to citation networks which are continuously growing with the publication of new papers.
Finally, \citet{scholtes2014causality} and \citet{vaccario2019mobility} identify temporal properties in the dynamics of real-world systems, which violate the path transitivity assumption.
These results raise concerns about how correctly model dynamic processes on networks, such as scientific credit diffusion and knowledge flow.

To address the problem introduced by the violation of the path transitivity assumption, \citet{scholtes2014causality, rosvall2014memory,lambiotte2019understanding} propose novel network models based on the path abstraction.
In this abstraction, instead of analyzing dyads, one looks at the time ordered path sequences between nodes.
Specifically, in citations, instead of concentrating on individual citation links, one should consider consecutive citations between articles to obtain \textit{citation paths}.
In our work, we use precisely this notion of citation paths to address the violation of path transitivity and its effect on journal rankings.

\section{Citation paths and the violation of path transitivity}
\label{sec:comp-paths-and-nets}

In citation data, we usually have a set of documents $\mathcal{D} = \{p_1, p_2,.., p_N\}$,
and a set of citation edges among them $\mathcal{E} = \{(p_2,p_1), (...), ... \}$
where $(p_j,p_i)$ represents a citation from document $p_j$ to $p_i$ with $i<j$.
Note that the subscript of documents represents their publication order.
So for example, $p_1$ is older than $p_2$ and $p_2$ is older of $p_3$ and so on and so forth.

\begin{figure}
	\footnotesize
	\center
	\includegraphics[width=1\textwidth]{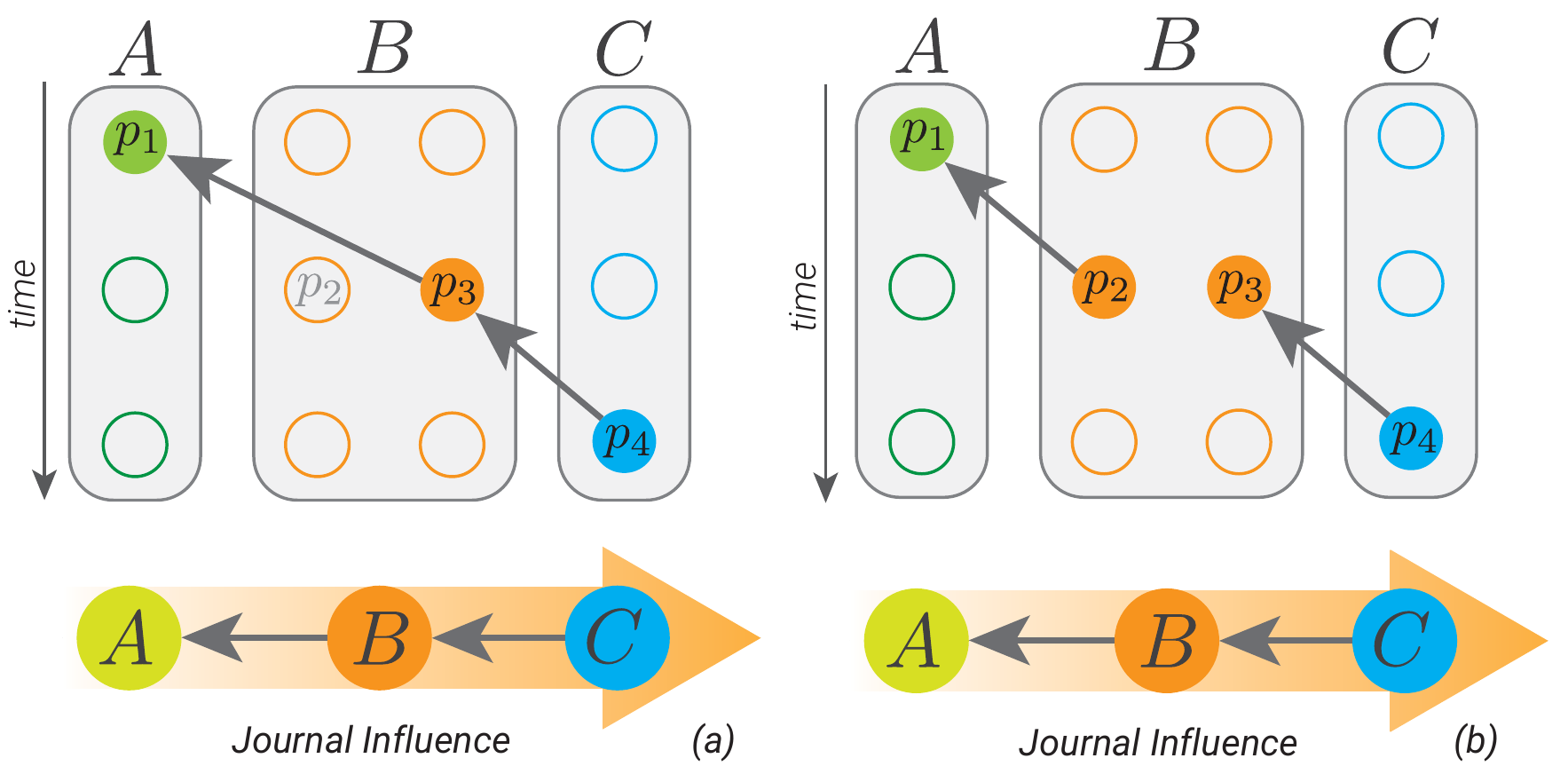}
	\caption{The citation projection from the paper to the journal level.
    In (a) we illustrate the case where journal $A$ may influence journal $C$ via journal $B$ through citation links.
    The citation network at journal level correctly captures this feature.
    However, in (b), $A$ cannot influence $C$ via $B$, because there is no uninterrupted citation path connecting the three journals.
    This fact is \textit{not captured} by the citation network at the journal level.
	}
	\label{fig:mixing}
\end{figure}

We restrict our attention to the case where the documents in $\mathcal{D}$ are scientific \textit{papers} published in \textit{journals}.
From the sets of papers, $\mathcal{D}$, and of citations, $\mathcal{C}$, we can build a citation network at the \textit{paper} level, where nodes are the papers and links are the citations.
One could argue that to investigate the citation network at the \textit{journal} level, we could define a new network
where (1) nodes are journals that contain the papers, and (2) links are the citations projected at the journal level.
Even though the first part is correct, the second step discards information required to quantify indirect inter journal influence.
To understand why this is the case, consider the example illustrated in Fig~\ref{fig:mixing}:

\begin{enumerate}
	\item[(a)] we have four papers $\mathcal{D} = \{p_1, p_2,p_3,p_4 \} $ and three journals $\mathcal{J}=\{A, B, C \} $.
		  The younger paper, $p_4$, belongs to journal $A$, the second and third papers, $p_2$ and $p_3$,
		  belong to journal $B$ and the older paper, $p_1$, belongs to journal $C$.
	      Additionally, we have the following citations $\mathcal{C} = \{ (p_4,p_3), (p_3, p_1)\} $.
	\item[(b)] we have the exact same setting as before, but we change one citation link: instead of $(p_3, p_1)$, we have $(p_2, p_1)$,
	i.e., $\mathcal{C}' = \{ (p_4,p_3), (p_2, p_1)\} $.
\end{enumerate}

In Figure~\ref{fig:mixing}, we build the citation network at the \textit{journal} level for both examples by aggregating and projecting the citations from the papers onto journals.
Here, we find that the citation networks at the {journal} level are the same.
However, the two citation networks at the \textit{paper} level are not the same, i.e., $\mathcal{C} \neq \mathcal{C}'$.
What do we miss by looking at the citation network at the {journal} level?
In the first case, Figure~\ref{fig:mixing}(a), we see that information, knowledge, and influence can propagate from journal $C$ to journal $A$ via journal $B$ thanks to the citation links.
In the second case (see top of figure Figure~\ref{fig:mixing}(b)), this is impossible as neither citations nor citation paths connect papers in the journals $A$ and $C$.
When looking at the citation network at the journal level, we cannot detect such a difference.

The standard projection of paper citations onto journals implies the existence of citation paths among journals that do not exist.
As illustrated in Figure~\ref{fig:mixing}~(b), the projection implies the existence of a citation link from $p_3$ to $p_2$ just because they are published in the same journal.
In other words, the projection introduces relations between journals that do not exist.
As mentioned in the Introduction, we refer to this problem as \textit{fictitious influence}.

\begin{figure}
    \centering
    \includegraphics[width=0.80\textwidth]{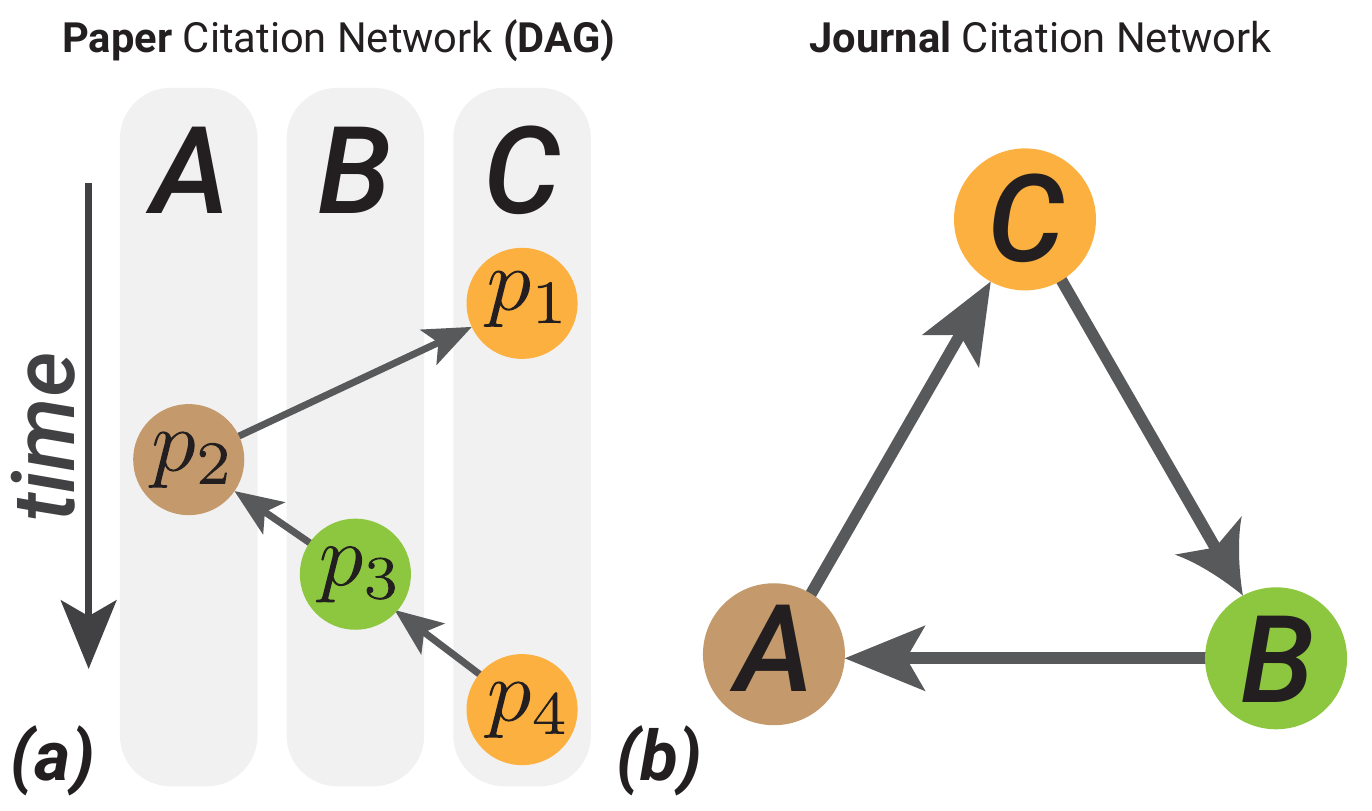}
    \caption{(a) An example of a paper citation network where four journal cite each other creating a direct acyclic graph.
    (b) The projection of the acyclic structure depicted in (a) onto journals: a cycle is created.}
    \label{fig:dag2cycle}
\end{figure}

On a higher level, one can understand the problem of fictitious influence by comparing the topology of paper and journal citation networks.
In paper citation networks, younger papers only cite older ones, and hence, we have \textit{direct acyclic graphs} (see Fig.~\ref{fig:dag2cycle}(a)).
On such topologies, we can define causal paths between papers: younger papers re-use knowledge and information of older papers, and cite them; in other words, older papers are a possible cause for the existence of younger ones.
This statement about a causal link does not have to be true as citation have different purposes \citep{bornmann2008citation}.
However, the acyclic topology of the citation network is a necessary (but not sufficient) condition for a causal connection to exist.
When one projects the citations at the journal level, one creates a network with many cycles, and breaks the possible causal structure captured by the acyclic topology (see Fig.~\ref{fig:dag2cycle}(b)).
In other words, one cannot define casual paths between journals, but only possible correlations between them.
Hence, when using non-local indicators on the journal citation network, one neglects the causal structure and introduces a fictitious influence between journals.
In the next section, we show that this fictitious influence affects journal rankings.

\section{An empirical investigation}\label{sec:recon-know-diff}

To quantify the importance of fictitious influence on journal rankings, we perform an empirical investigation.
Precisely, we construct the citation network for papers and its projection at the journal level.
Then, we will use these two networks to derive two journal rankings based on PageRank.
We compute the first ranking on the journal citation network and the fictitious influence will affect this ranking.
Then, we compute the second ranking using citation paths extracted on the paper citation networks, and the fictitious influence will not be affect this second ranking. 
A stark discrepancy in the two rankings would indicate that fictitious influence is not innocuous.
We choose PageRank since it is a prototypical non-local indicator used to rank journals \citep{guerrero2012further} in addition to websites \citep{brin1998anatomy}.

\subsection{Data}
\begin{figure}
	\footnotesize
	\begin{center}
        (a)\includegraphics[width=0.46\textwidth]{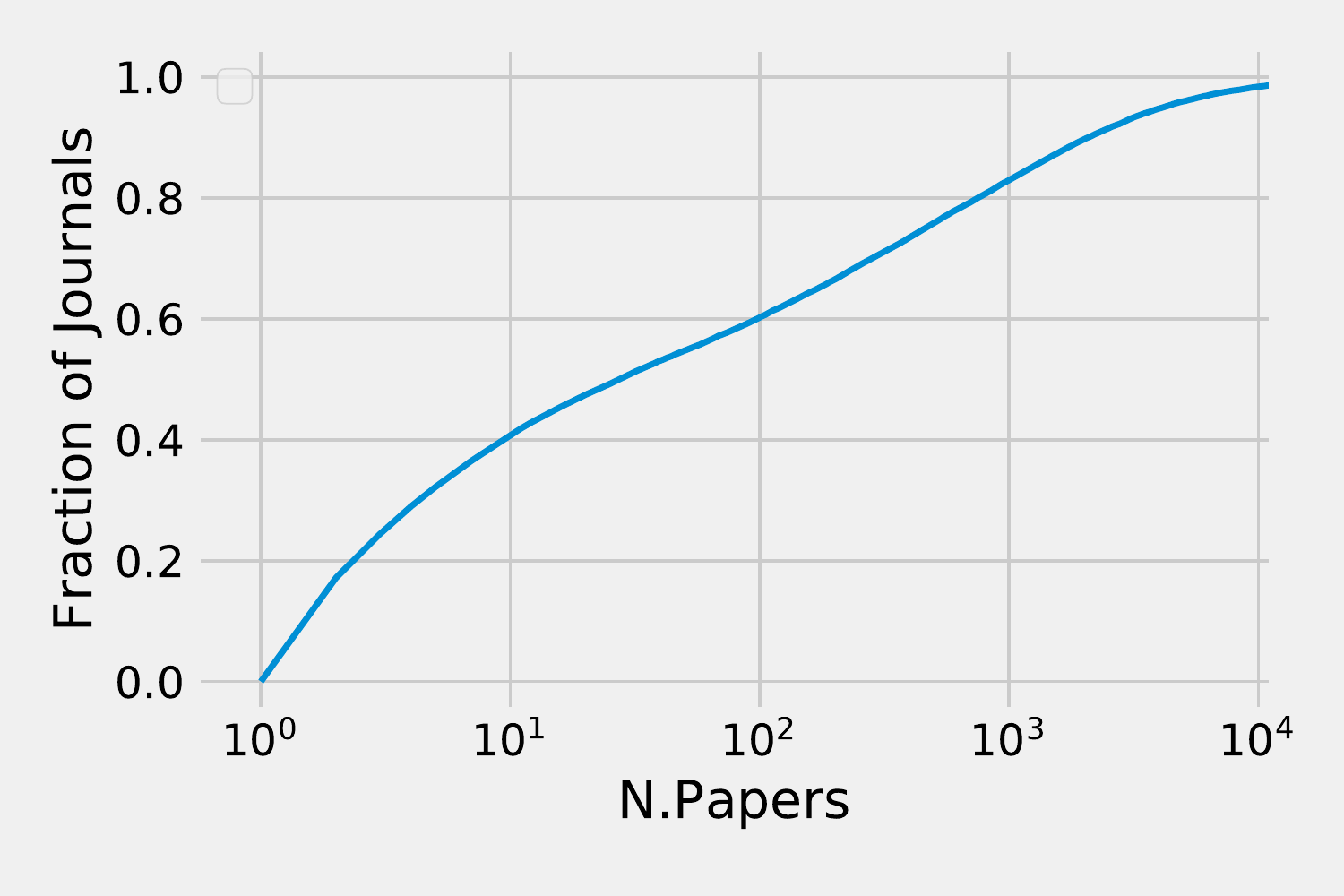}
        (b)\includegraphics[width=0.46\textwidth]{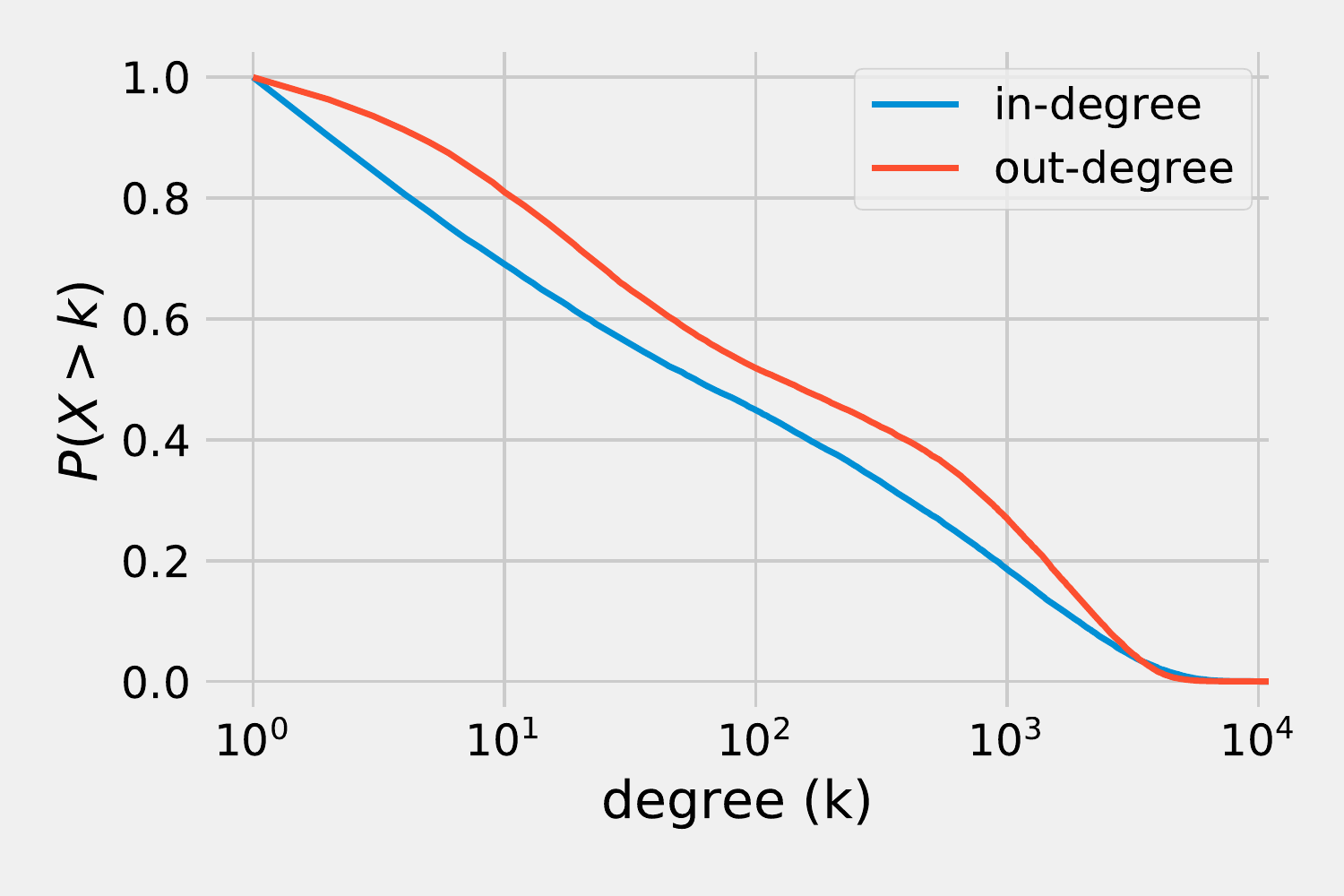}
	\end{center}
	\caption{Number of papers and citations per journal.
		(a) Fraction of journals with at least a given number of papers.
		(b) Counter cumulative distribution of incoming (in-degree, blue) and outgoing citations (out-degree blue) per journal.}
	\label{fig:journals_counts}
\end{figure}

We use citation data from MEDLINE obtained from the Torvik Group by combining various public available sources, including the MAG, AMiner and PumbedCentral~\url{http://abel.lis.illinois.edu/}
\footnote{Access to this data was obtained by getting in contact with the Torvik group.}.
The data contains detailed information about $26\,759\,399$ papers published between 1940 and 2016 with more than 460 Mio citations.
To link the papers to the journals, we use a dump of the Pubmed\url{https://www.nlm.nih.gov/databases/download/pubmed_medline.html}.
We find that these paper belong to $24\,135$ journals and have $323\,356\,788$ citations.
Note that more than 50\% of the journals have at least 20 papers, 50 incoming citations, and 100 outgoing citations (see Figure~\ref{fig:journals_counts}).

\subsection{Methods}\label{sec:emp-diff-proj-links-and-paths}

To rank journals according to PageRank, the standard approach is to project citation from papers onto journals to obtain a journal citation network.
On such a network, it is then possible to compute PageRank centrality according to:
\begin{equation}
	\textrm{PR} = \left( \frac{1-d}{n} E + d T \right) \textrm{PR}
\end{equation}
where $\textrm{PR}$ is the vector containing the PageRank scores of the nodes, $d$ the damping factor\footnote{We choose $d=0.5$ as proposed by~\citep{chen2007finding}.}, $E$ is a $n\times n$ matrix of 1s, and $T$ is the transition matrix of the journal citation network \citep{brin1998anatomy}.
As discussed in the previous section, this standard approach introduces fictitious influence between journals.

To understand how to avoid the fictitious influence when computing PageRank scores, let us recall the dynamic process captured by this centrality.
In this process, we have a random walker that is placed on a node.
From this node, the walker can either follow a link, or ``teleport'' to a random node in the network.
Then, the PageRank score of a node is its visitation probability, i.e., how likely is to find the walker on that node.

The simplest way to address the fictitious influence problem is to unfold the random walk on the paper citation network, instead of the journal citation network.
Indeed, on the paper citation network, the random walker can \textit{only} follow the empirical citation paths.
To rank journals according to PageRank computed on these paths, we i) place the random walker on a journal, ii) move the walker on a random paper belonging to the journal, iii) let the walker follow the citation paths (i.e., on the paper citation network), or ``teleport'' to a random journal
\footnote{Note that the teleportation occurs at the journal level as we want to capture journal importance using PageRank. After teleporting to random journal we are back to step i). To operationalize this PageRank algorithm is sufficient to compute the PageRank on the paper network with a personalized starting vector.}.
Depending on the visitation probability of papers, we obtain the paper PageRank scores.
By summing the PageRank scores of papers belonging to the same journal, we obtain the overall journal PageRank scores $\mathrm{PR}_{\mathcal{C}}$.

\subsection{Results}

\begin{table}
	\centering
	\footnotesize
\begin{tabular}{cccl}
    \toprule
    $\textrm{PR}$-rank	& $\textrm{PR}_{\mathcal{C}}$-rank	& Change 	& Journal Name \\
    \midrule
     1 	&  4 	&  \color{eth_red}{-3$\bm{\downarrow}$}	    &  Science \\
     2 	&  2 	&              {$\bm{=}$} 	                &  Proc. Natl. Acad. Sci. U.S.A. \\
     3 	&  3 	&              {$\bm{=}$} 	                &  Nature \\
     4 	&  1 	&  \color{eth_green}{+3$\bm{\uparrow}$}     &  J. Biol. Chem. \\
     5 	&  5 	&               {$\bm{=}$} 	                &  N. Engl. J. Med. \\
     6 	&  6 	&               {$\bm{=}$} 	                &  Lancet \\
     7 	&  9    &  \color{eth_red}{-2$\bm{\downarrow}$} 	&  JAMA \\
     8 	&  10 	&  \color{eth_red}{-2$\bm{\downarrow}$} 	&  Cell \\
     9 	&  7 	&  \color{eth_green}{+2$\bm{\uparrow}$}     &  Circulation \\
     10 &  14 	&  \color{eth_red}{-4$\bm{\downarrow}$}	    &  J. Clin. Invest. \\
     11 &  13 	&  \color{eth_red}{-2$\bm{\downarrow}$} 	&  J. Immunol. \\
     12 &  12 	&               {$\bm{=}$} 	                &  Cancer Res. \\
     13 &  15 	&  \color{eth_red}{-2$\bm{\downarrow}$}     &  Blood \\
     14 &  19 	&  \color{eth_red}{-5$\bm{\downarrow}$}     &  BMJ \\
     15 &  20 	&  \color{eth_red}{-5$\bm{\downarrow}$}     &  Nucleic Acids Res. \\
     16 &  34 	&  \color{eth_red}{-8$\bm{\downarrow}$}     &  J. Neurosci. \\
     17 &  36   &  \color{eth_red}{-9$\bm{\downarrow}$}     &  Pediatrics \\
     18 &  106 	&  \color{eth_red}{-88$\bm{\downarrow}$}  &  Am J Public Health \\
     19 &  22 	&  \color{eth_red}{-3$\bm{\downarrow}$}     &  J. Exp. Med. \\
     20 &  29 	&  \color{eth_red}{-9$\bm{\downarrow}$}     &  Ann. Intern. Med. \\
     \bottomrule
\end{tabular}
\caption{Top-20 journals according to $\textrm{PR}$.
The first column ($\textrm{PR}$) contains the rank of the journal as computed on the journal citation network.
The second column ($\textrm{PR}_{\mathcal{C}}$) contains the rank of the journal as computed using the citation paths.
The Change column contains an arrow pointing downwards when the journal loses positions in the $\textrm{PR}_{\mathcal{C}}$-ranking, an upward arrow if the journal gains positions, and an equal sign if the rank is the same.
}
\label{tab:top_20_journal_pageRank_std}
\end{table}

\begin{figure}[h]
    \centering
    \includegraphics[width=1\textwidth]{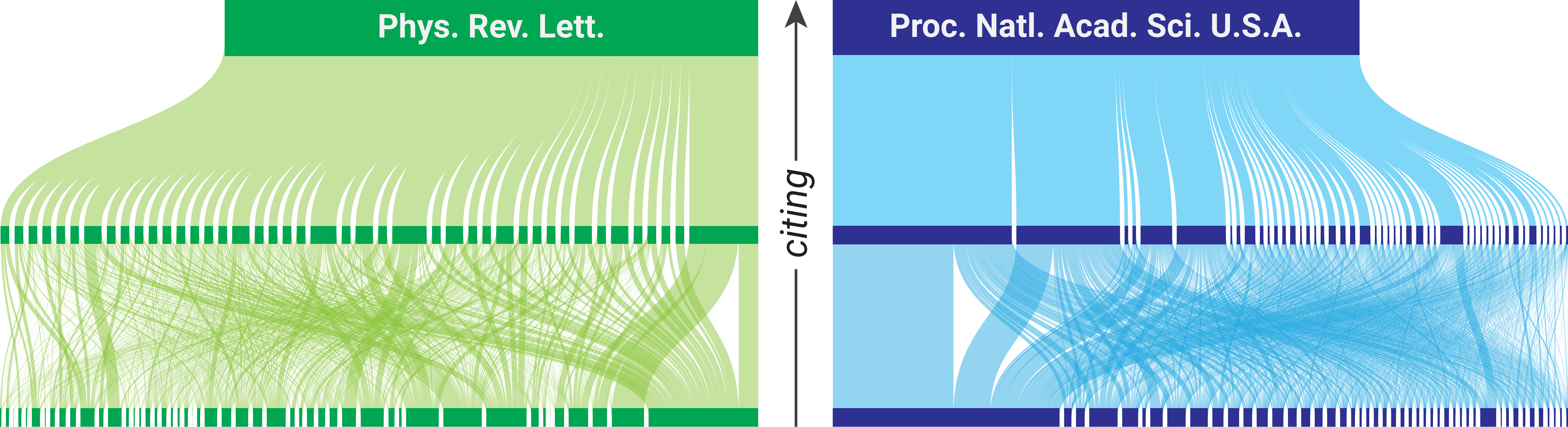}
    \caption{Citation paths of length two arriving at papers published in ``Phys. Rev. Lett.'' (left) and ``Proc. Natl. Acad. Sci. U.S.A.'' (right) from the 45 top-ranked journals (according to the standard PageRank measure). 
    In this alluvial diagram the top node represents the focal final nodes: ``Phys. Rev. Lett.'' (left) and ``Proc. Natl. Acad. Sci. U.S.A.''. %
    The second and third row show the proportion of citations that pass through the 45 top-ranked journals to arrive to the focal nodes.}
    \label{fig:prl_pnas_example}
\end{figure}

The number of unique citation paths of length two observed in the dataset is 1\,095\,968\,097.
In Figure~\ref{fig:prl_pnas_example} we show an example of the extracted citations paths reaching PNAS and PRL in two steps.
In this representation we see the 45 journals citing most often the respective focal journals.
The figures shows the variety and distribution of citations paths leading to PRL and PNAS.
When projecting the paper citations at journal level, the number of implied paths is 340\,997\,180\,016.
This means that by projecting the citations we introduce more than 300 Billion paths which are never observed in the data.
These are the paths which may give rise to fictitious influence.
Note that if we consider even longer paths, i.e., more than 2, the problem becomes even more pronounced.

In Table~\ref{tab:top_20_journal_pageRank_std},
we report the rankings of the top-20 journals according to PageRank computed with the standard network approach ($\mathrm{PR}$).
Additionally, we also report the rank position of these top-20 journals according to PageRank computed using the empirical citation paths ($\mathrm{PR}_{\mathcal{C}}$).
In this table, we see that several journals change their position within the ranking.
For example, we find that the ranking of journals like ``Proc. Natl. Acad. Sci. U.S.A.'', ``Nature'' and ``Lancet'' are not affected.
In contrast, journals like ``Science'', ``J. Neurosci.'' and ``Am J Public Health'' lose several positions.
One extreme example is the ``Am J Public Health'', which moves from the $18^\textrm{th}$ to $106^\textrm{th}$ position in the ranking.
For other journals, like ``PJ. Biol. Chem.'', we see an improvement in their ranking position.
In the Appendix, we report the top-20 journals according to $\mathrm{PR}_{\mathcal{C}}$ for completeness (see  Table~\ref{tab:top_20_journal_pageRank_pers}).

\begin{figure}
	\footnotesize
    \begin{center}
        \includegraphics[width=0.55\textwidth]{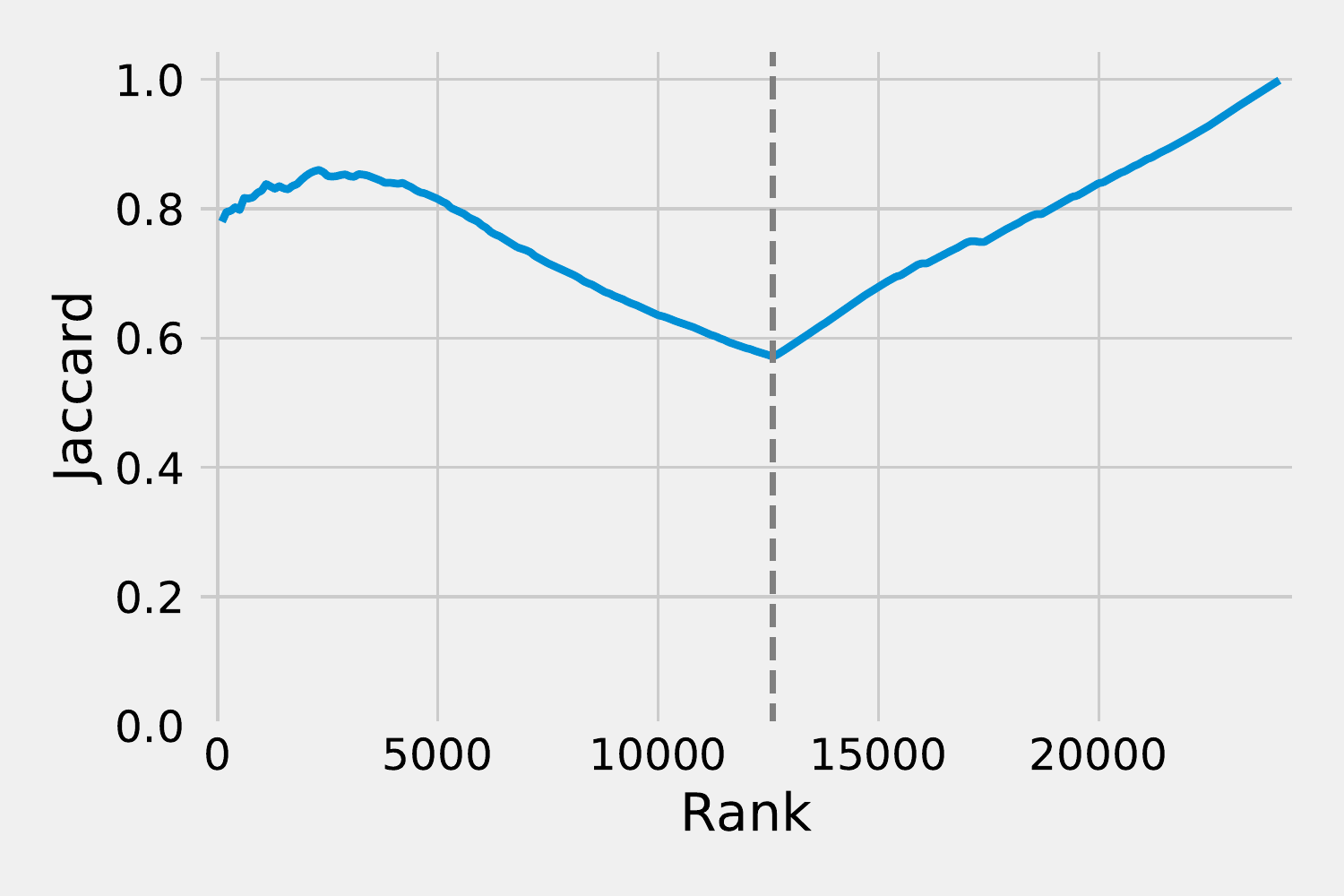}
    \end{center}
    \caption{We plot the overlap between the two rankings, i.e., we consider the top-$k$ ranked journals according to $\textrm{PR}$ and compute the intersection with the top-$k$ ranked journals according to $\textrm{PR}_{\mathcal{C}}$.}
    \label{fig:overlap}
\end{figure}

\begin{figure}
	\footnotesize
    \begin{center}
        \includegraphics[width=0.55\textwidth]{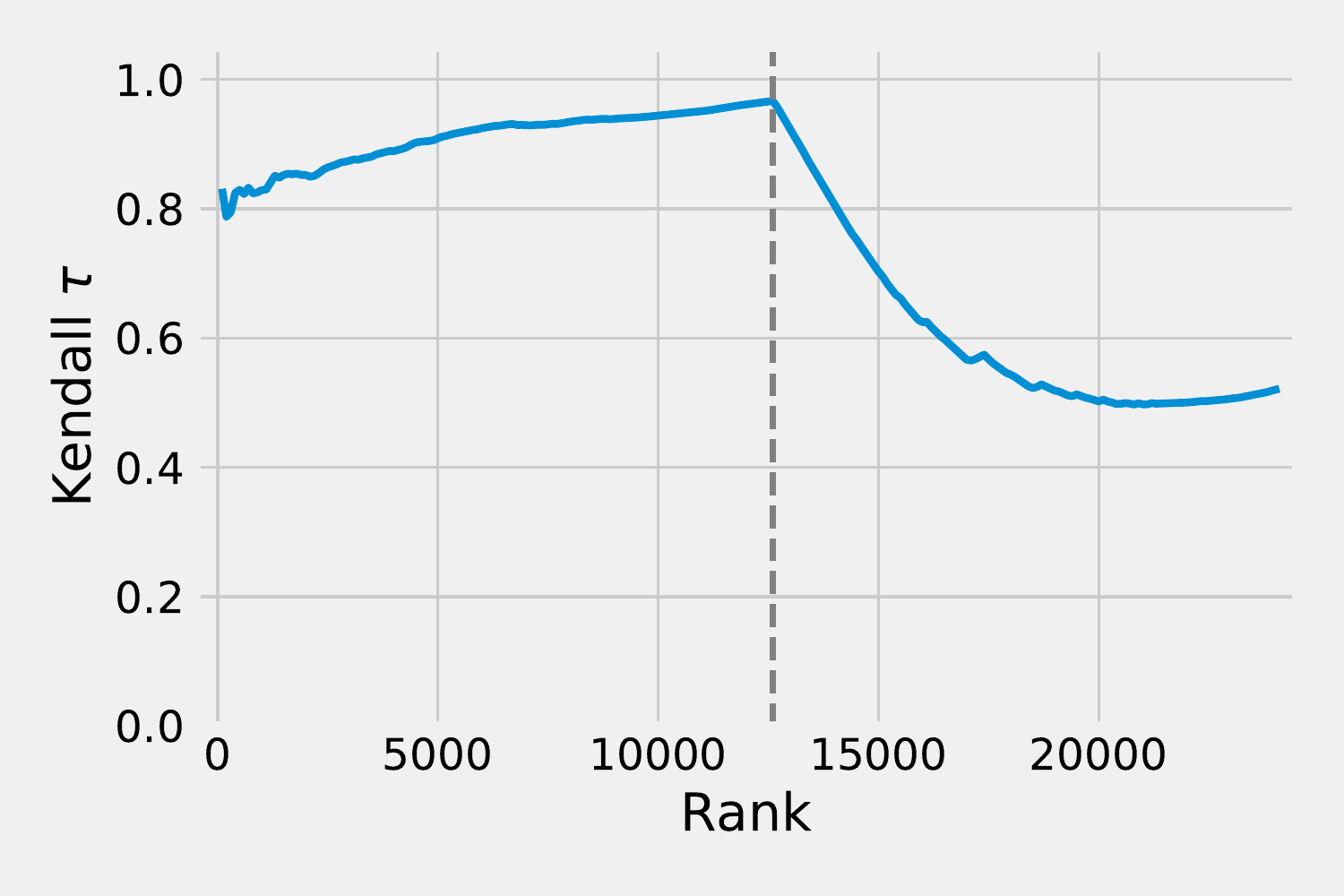}
	\end{center}
	\caption{Comparison between the two ranking.
        We plot the Kendall's $\tau$ coefficient between the top-$k$ ranked journal according to the PageRank computed on the journal citation network and the corresponding relative rankings of these $k$ journals of journals according to the PageRank computed using the empirical citation paths.
        }
	\label{fig:kt}
\end{figure}

To quantify the difference between the two rankings, we first compute the overlap between the rankings.
Precisely, we calculate the Jaccard similarity between two sets of journals listed among the top-$k$ journals according to the two approaches.
In Fig.~\ref{fig:overlap}, we report this similarity for different values of $k$.
We see that for small values of $k$, i.e., when considering the top positions, we have about 80\% overlap indicating that the rankings share the same 80\% of journals in these top positions.
However, when comparing larger fraction of the rankings, the intersection decreases to 60\%.
In other words, almost half of the journals listed in the two rankings are different.
This indicates that the two rankings are extremely different.
For larger values, the intersection increases linearly to the value 1. This result is expected as the complete rankings contain the same journals and their similarity is trivially 1. %

To further quantify the difference between the rankings coming from $\mathrm{PR}$ and $\mathrm{PR}_{\mathcal{C}}$, we compute the Kendall $\tau$ coefficient (KT)\citep{kendall1945treatment}.
When considering the full ranking, we obtain a low value around $0.5$.
Similar to before, we also compute the KT-coefficient by considering the top-$k$ journals according $\mathrm{PR}$ for different values of $k$.
In Fig.~\ref{fig:kt}, we report how the KT-coefficient changes with $k$.
We find that it increases when considering the first $\approx12\,500$ ranked journal, and then we have a sharp decrease.
First note that the increase of the KT-coefficient does not imply that the ranking are similar as only less than 60\% of the journals are the same.
It only means that the relative positions of these 60\% of journals are correlated.

Second the sharp decrease of the KT-coefficient marks the point where $\mathrm{PR}$ fails in ranking the journals.
Indeed, $\mathrm{PR}$ assigns to many journals the same scores for position lower than $12\,500$.
In contrast, $\mathrm{PR}_{\mathcal{C}}$ that uses the empirical citation paths assigns unique PageRank scores also to these less central journals.
Note that to rank these journals, $\mathrm{PR}_{\mathcal{C}}$ relies on fewer assumptions, i.e., we have relaxed the transitivity assumption.

The rankings created with and without correcting for fictitious influence are substantially different.
In other words, the discrepancy in the rankings indicates that computing the network measure on the journal citation network yields wrong and possibly misleading results.

\section{Discussion}\label{sec:chap4-con}

Increasing attention has been given to data to guide science and research policy~\cite{hicks2015bibliometrics}.
This usage has produced the need to develop new and more sophisticated measures to quantify scientific performance.
In particular, several measures have been constructed by combining bibliometric and network methods.
In this work, focusing on measures for journal impact, we have shown how a naive combination of these methods may lead to misleading or even wrong results.
Specifically, we have argued that a standard projection of paper citations onto journals may introduce nonexistent relations, which we call fictitious influence.

\textit{First} we have explained how fictitious influence arises from the transitivity assumption, which is a common and central assumption in many standard network methods.
In particular, we have identified two ways how fictitious influence may arise: i) the time and ii) the journal aggregation of citation links.
By time aggregating citations, one loses the ordering in which citations occurs between journals.
By aggregating citations inside journals, one mixes the incoming and outgoing citations of papers belonging to the same journal.
These aggregations introduce relations between journals that do not respect the empirical citation patterns among papers.

\textit{Second} we have shown that the fictitious influence is not an innocuous effect when computing rankings of journals.
To do this, we have used real world citation data coming from MEDLINE, the largest open-access bibliometric dataset in the life science.
With this data, we have first computed the number of paths of length 2 on the paper citation and on the journal citation network.
In the former represent the empirically observed paths, whereas the latter represent implied paths after projecting paper citations onto journals.
We find that only 0.3\% of the implied citation paths are actually present in the dataset.
This discrepancy highlights that the projection introduce a large number of wrong citation paths, allowing for fictitious influence.
Then we have computed two journal rankings using the standard journal citation network, and the paper citation network.
On the former network, we have computed the PageRank scores of journals that are biased from the fictitious influence.
On the latter network, we have computed the unbiased PageRank scores.
Among the top 2500 journals, we have found that the overlap between the rankings is relatively high ($\approx0.85$) and a low Kendall's $\tau$ 0.70.
These results indicates that even though the same journals belong to the top of the rankings, they occupy different positions.
When considering the top 12500 journals, we have found that the overlap between the rankings decreases to approximately $0.60$, but the Kendall's $\tau$ increases.
This indicates that the two rankings become extremely different as they share less than 60\% of the journals, but the relative positions of these journals is consistent across rankings.
Overall, our results indicate the the fictitious influence has a strong effect when using PageRank to rank journals .

To overcome the problem of fictitious influence, one could argue that higher-order networks are a possible solutions.
On one hand, these network model could help because centrality measures computed on them correlate to the one coming from the original sequence data~\citep{scholtes2017kdd}.
On the other hand, they rely on the assumption that there are temporal correlations in the data allowing to summarize it.
For an overview of application pf these model to various data see~\citep{lambiotte2019understanding}.
Hence, the viability of these methods will depend on the specific research questions addressed.

This work has the following primary limitations.
We used only citation data from MEDLINE, a databases with primary focus on bibliometric information in the life sciences.
Hence, we have analyzed a biased sample of bibliographic data. 
This limits the reliability of the obtained rankings. 
However, the discrepancies found between the rankings highlight the fundamental problem of fictitious influence. 
The second limitation is that we only considered one possible non-local indicator, PageRank. 
There are many other non-local network indicators, and on each of them, the effect of fictitious influence could be different. 
To address these limitations, future works can replicate our analysis on a more complete citation dataset and by considering other non-local indicators.

To conclude, we have shown that journal rankings based on non-local journal indicators may be wrong.
This problem arises because a naive projection of paper citations onto journals introduces fictitious relations.
To overcome this problem, we have proposed to adopt a path perspective.
With this work, we highlighted the shortcomings of the standard network approach to create journals rankings.
Also, we provide a new perspective to use citation analysis at the journal level to support research evaluators and administrators in the challenging tasks of assessing scientific performance.

\subsection*{Acknowledgements}
We thank Frank Schweitzer for helpful discussions. Also, we thank Ingo Scholtes for his many critiques and suggestions, which improved the manuscript.

\subsection*{Conflict of interest}

The authors declare that they have no conflict of interest.

\bibliographystyle{spbasic}      %

\bibliography{all_references}

\appendix
\clearpage
\section{top-20 journals according to $\mathrm{PR}_{\mathcal{C}}$}

\begin{table}[!h]
	\centering
	\footnotesize
\begin{tabular}{cccl}
    \toprule
    $\textrm{PR}$-rank	& $\textrm{PR}_{\mathcal{C}}$-rank& Change 	& Journal Name \\
    \midrule
     4 	&  1 	&   \color{eth_green}{+3$\bm{\uparrow}$} 	    &  J. Biol. Chem. \\
     2 	&  2 	&              {$\bm{=}$} 	                &  Proc. Natl. Acad. Sci. U.S.A. \\
     3 	&  3 	&              {$\bm{=}$} 	                &  Nature \\
     1 	&  4 	&  \color{eth_red}{-3$\bm{\downarrow}$} 	    &  Science \\
     5 	&  5 	&             {$\bm{=}$} 	                &  N. Engl. J. Med. \\
     6 	&  6 	&              {$\bm{=}$} 	                &  Lancet \\
     9 	&  7    &  \color{eth_red}{-2$\bm{\downarrow}$} 	&  Circulation \\
     27 &  8 	&  \color{eth_green}{+19$\bm{\uparrow}$} 	&  Phys. Rev. Lett. \\
     7 	&  9 	&  \color{eth_red}{-2$\bm{\downarrow}$} 	&  JAMA \\
     8  &  10 	&  \color{eth_red}{-2$\bm{\downarrow}$} 	&  Cell \\
     22 &  11 	&  \color{eth_green}{+11$\bm{\uparrow}$} 	&  Biochim. Biophys. Acta \\
     12 &  12 	&              {$\bm{=}$} 	                &  Cancer Res. \\
     11 &  13	&  \color{eth_red}{-2$\bm{\downarrow}$} 	&  J. Immunol. \\
     10 &  14 	&  \color{eth_red}{-4$\bm{\downarrow}$} 	&  J. Clin. Invest. \\
     13 &  15 	&  \color{eth_red}{-2$\bm{\downarrow}$} 	&  Blood \\
     31 &  16   &  \color{eth_green}{-15$\bm{\uparrow}$} 	&  Biochem. J. \\
     24 &  17 	&  \color{eth_green}{+7$\bm{\uparrow}$} 	&  Biochemistry \\
     25 &  18 	&  \color{eth_green}{+7$\bm{\uparrow}$} 	&  Cancer 		 \\
     14 &  19 	&  \color{eth_red}{-5$\bm{\downarrow}$} 	&  BMJ \\
     15 &  20 	&  \color{eth_red}{-5$\bm{\downarrow}$} 	&  Nucleic Acids Res. \\
     \bottomrule
\end{tabular}
\caption{Top-20 journals according to $\textrm{PR}_{\mathcal{C}}$.
The first column ($\textrm{PR}$) contains the rank of the journal computed on the journal citation network.
The second column ($\textrm{PR}_{\mathcal{C}}$) contains the rank of the journal computed using the citation paths.
The Change column contains an arrow pointing downwards when the journal loses positions in the $\textrm{PR}_{\mathcal{C}}$-ranking, an upward arrow if the journal gains positions, and an equal sign if the rank is the same.
}
\label{tab:top_20_journal_pageRank_pers}
\end{table}

\end{document}